\documentclass[twocolumn,showpacs,superscriptaddress,showpacs,floatfix]{revtex4}

\usepackage{epsfig,color}
\usepackage{graphicx}
\usepackage{dcolumn}
\usepackage{bm}
\usepackage{setspace}

\begin{document}

\preprint{APS/123-QED}

\title{Connecting the Reentrant Insulating Phase and the Zero Field Metal-Insulator Transition in a 2D Hole System}
\author{R. L. J. Qiu}
\author{X. P. A. Gao}
\email{xuan.gao@case.edu}
\affiliation{Department of Physics, Case Western Reserve University, Cleveland, OH 44106}
\author{L. N. Pfeiffer}
\author{K. W. West}
\affiliation{Department of Electrical Engineering, Princeton University, Princeton, NJ 08544}

\date{\today}

\begin{abstract}

We present the transport and capacitance measurements of 10nm wide
GaAs quantum wells with hole densities around the critical point
of the 2D metal-insulator transition (critical density $p_c$ down
to 0.8$\times$10$^{10}$/cm$^2$, $r_s$$\sim$36). For metallic hole
density
$p_c$$\textless$$p$$\textless$$p_c$+0.15$\times$10$^{10}$/cm$^2$,
a reentrant insulating phase (RIP) is observed between the $\nu$=1
quantum Hall state and the zero field metallic state and is
attributed to the formation of pinned Wigner crystal. Through
studying the evolution of the RIP versus 2D hole density by
transport and capacitance experiments, we show that the RIP is
incompressible and continuously connected to the zero field
insulator, suggesting a similar origin for these two phases.

\end{abstract}
\maketitle


Strongly correlated electron systems have offered numerous
interesting topics for mankind to explore, e.g. Fermi liquid,
Wigner crystal (WC), fractional quantum Hall (QH) effect
\cite{FQHE}, and high temperature
superconductivity\cite{HighTcSC}. In the two dimensional (2D)
case, a possible phase transition between metal and insulator,
which contradicts the scaling theory of localization for
non-interacting 2D fermions\cite{AndersonWL}, was discovered by
Kravchenko {\it et al.} \cite{MIT}. The origin of this
metal-insulator transition (MIT) in strongly correlated 2D
electron or hole systems has been of great
interest\cite{mitreview,mitreview2}. A relevant question is how
the zero field ($B$=0) MIT is related to the 2D QH states in the
presence of a perpendicular magnetic field ($B_{\perp}$). In early
experiments, the phase diagram maps against density $p$ and
magnetic field $B_{\perp}$ obtained from both
transport\cite{DCTsuiPhasediagram} and thermodynamic
compressibility\cite{HWJiangCvsBT} measurements suggested that
$B_{\perp}$ transforms the $B$=0 insulator into the QH liquid,
somewhat similar to the conventional Anderson insulator-QH
transition\cite{HWJiangPRL1993}. Yet this $B$=0 insulator to QH
transition moves to lower field and eventually terminates at $p_c$
of the $B$=0 MIT as carrier density $p$ is increased
\cite{DCTsuiPhasediagram}. Besides the zero field MIT and $B$=0
insulator to $\nu$=1 QH transition, 2D electron or hole systems
are known to exhibit rich transitions between the so called high
field insulating phases (HFIP)\cite{HFIP}, reentrant insulating
phases(RIPs)
\cite{VJGoldmann-GaAsreentrant,HWJiange-GaAsreentrant,HWJiangp-GaAsreentrant,MBSantosp-GaAsreentrant},
and the fractional QH states at $\nu<$1. All the HFIPs and RIPs
among fractional QH states in GaAs/AlGaAs systems are widely
believed to be caused by the forming/melting of Wigner
crystal\cite{VJGoldmann-GaAsreentrant,HWJiange-GaAsreentrant,HWJiangp-GaAsreentrant,MBSantosp-GaAsreentrant,lowfnoise,WCbook,p-GaAsreentrant2005,naturephysics2008}
due to the strong Coulomb interaction. Moreover, the observations
of pinning mode of the WC oscillating in the disorder potential in
microwave transmission
experiments\cite{microwavetransmission,microwavetransmission2002,microwavetransmission2004,microwavetransmission2010}
offer more direct evidence for the WC interpretation.

On the theory side, quantum Monte Carlo calculations
\cite{MonteCarlo,WCtheory,WCtheorytwo,XZhuMonteCarlo} had located
that the liquid-Wigner crystal transition should happen around a
critical value of $r_{s}\sim$37 at $B$=0; the simulations even
determined the phase boundary\cite{XZhuMonteCarlo} starts from
$\nu$$\sim$1/6.5 and moves to higher $\nu$ when  $r_s$ increases
(density decreases). The results, together with other variational
studies\cite{FZ-diagonalization}, supported the interpretation of
the observed HFIPs and RIPs as pinned Wigner solids. However, the
validity of the phase diagram proposed by Zhu and Louie
(Ref.\onlinecite{XZhuMonteCarlo}(b)) remains
unsettled\cite{MonteCarlo2000+,WCvsCF}, especially in the range
from $\nu$=1 to the $B$=0 limit, since a reentrant WC that was
expected at $\nu$$\textgreater$1 has not been observed
experimentally. Therefore, the relation between those HFIPs, RIPs
and zero field insulating phase in 2D systems with large $r_s$ is
still unclear.


In this Letter, we report the first observation of a reentrant
insulating phase between the $\nu$=1 quantum Hall state and $B$=0
metallic state in dilute 2D GaAs/AlGaAs hole system with stronger
quantum confinement (narrower quantum well). We suggest that this
RIP is a pinned Wigner crystal since it is \emph{incompressible}
\cite{HWJiangCvsBT,WCcompressibility} and its peak resistance
$R_{RIP}$$\propto$exp($\Delta_{RIP}$$/2T$) ($\Delta_{RIP}$ up to
0.52K at $p$=0.55$\times$10$^{10}$/cm$^2$ for our highest mobility
sample), resembling the previously observed low $\nu$ RIPs in GaAs
dilute 2D
electron\cite{VJGoldmann-GaAsreentrant,HWJiange-GaAsreentrant} or
hole\cite{HWJiangp-GaAsreentrant,MBSantosp-GaAsreentrant,p-GaAsreentrant2005}
systems. Our transport and compressibility phase diagrams at
$T$=70mK indicate a unified phase boundary linking zero field MIT
with the RIP between $\nu$=1 and $\nu$=2 QH states and the HFIP,
consistent with the qualitative phase diagram proposed by
Ref.\onlinecite{XZhuMonteCarlo}(b). The results presented here
suggest that the zero-field MIT is a liquid-WC transition.


Our transport measurements were performed down to 50mK on four
high mobility 10nm wide p-GaAs quantum well (QW) samples (5A, 5B,
11C, and 11D) with low 2D hole density. Thermodynamic
compressibility measurement was done on QW 11D by measuring the
capacitance between the 2D channel and top gate. All of the four
GaAs QWs are similar to those used in our previous
studies\cite{Gaoprl,Gaoprl2,GaoparallelB,RichardandGao}, which
were grown on (311)A GaAs wafer using Al$_{.1}$Ga$_{.9}$As
barriers and Si delta-doping layers placed symmetrically at a
distance of 195 nm away from the 10nm GaAs QW. QW 5A and 5B were
taken from wafer 5-24-01.1, and sample 11C and 11D were taken from
another wafer 11-29-10.1. All the samples have density $p\sim$1.3
(the unit for $p$ is 10$^{10}$/cm$^2$ throughout this Letter) from
doping. Without gating, QW 5A and 5B have a mobility $\simeq$
5$\times$10$^{5}$cm$^2$/Vs, while QW 11C and 11D's mobility
$\simeq$2$\times$10$^{5}$cm$^2$/Vs. The back gates used to tune
the hole density are approximately 150$\mu$m away from the well
for QW 5A and 5B (300$\mu$m spacing for 11C), to avoid the
screening of the Coulomb interaction between holes by the back
gate. Sample 5A and 5B were fabricated into Hall bars with
approximate dimensions: 5A-2.5mm$\times$9mm; 5B-2mm$\times$6mm,
and six diffused In(1$\%$Zn) contacts; sample 11C is a square one
(5mm$\times$5mm) with diffused In(1$\%$Zn) contacts in Van de Pauw
configuration. The measurement current was along the high mobility
[$\underline 2$33] direction and stayed low so the heating power
is less than 3f Watt/cm$^2$ to prevent overheating\cite{GaoPhonon}
the holes. The compressibility measurement sample QW 11D has a
size of 4.5mm$\times$5mm, with four diffused In(1$\%$Zn) contacts
in Van de Pauw configuration and coated with Ti/Au top gate, which
is 400nm away from the well. Capacitance was measured by applying
a low frequency ($f$=3-13 Hz) excitation voltage \emph{$V_{ex}$}
(typically 200 $\mu$V) to the top gate and monitoring the
$90^{\circ}$ out of phase current \emph{$I_y$}, obtaining the
capacitance $C$=$I_y$/2$\pi f V_{ex}$. A DC voltage \emph{$V_g$}
was superimposed to tune the density via top gate. The measured
capacitance can be viewed as due to two capacitors in series:
$1/C=1/C_0+1/C_q$, where $C_0$ is geometric parallel plate
capacitance between the top gate and the 2D hole system, and $C_q$
is the quantum capacitance of charging the 2D hole system. The 2D
hole system's quantum capacitance \emph{$C_q$} is proportional to
the compressibility $\kappa$: \emph{$C_q$}=$e^2$d\emph{p}/d$\mu$
=$e^2$$\emph{p}^2$$\kappa$ (here $\mu$ is the chemical potential,
d\emph{p}/d$\mu$ is the thermodynamic density of states.).


\begin{figure}[btph]
\centerline{\epsfig{file=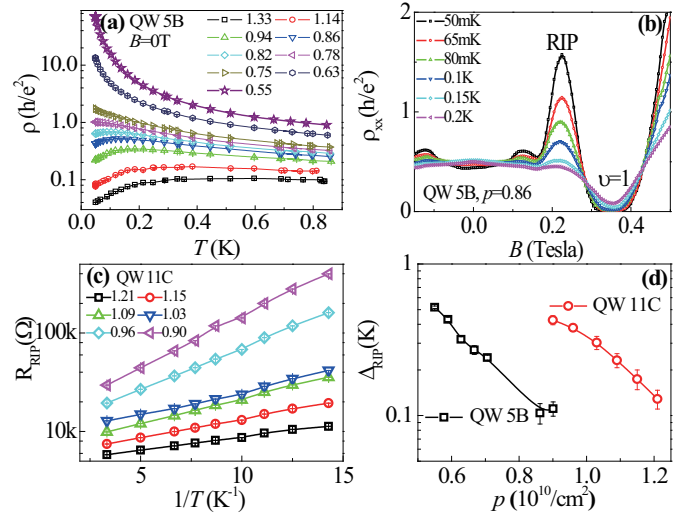,width=9.5cm}} \caption{(color
online) (a)Zero field metal-insulator transition (MIT) with a
critical density ($p_c$$\sim$0.8$\times$10$^{10}$/cm$^2$) in a
10nm wide GaAs QW sample 5B.(b)Magneto-resistivity of sample 5B
when $p$=0.86 at the temperatures of 50mK, 65mK, 80mK, 0.1K,
0.15K, and 0.2K. The reentrant insulating phase (RIP) resides
between the $\nu$=1 QH state and the zero field metallic
state.(c)Arrhenius plot of the resistance at the peak of the RIP
from sample 11C at densities of 0.9, 0.96, 1.03, 1.09, 1.15 and
1.21.(d)Fitted thermal activation gap $\Delta_{RIP}$ Vs. hole
densities in sample 5B and 11C.} \label{fig1}
\end{figure}


Fig.\ref{fig1} is shown to establish the existence of
metal-insulator  transition and reentrant insulating behavior at
$\nu>$1 in our low-density 2D hole system. With the tuning of the
back gate\cite{GaoHall}, we were able to do transport measurement
in a density range of 0.45-2.2, corresponding to a
$r_s$$\sim$22-48 (using effective hole mass
$m^*$=0.3$m_e$\cite{mass}). Figure \ref{fig1}(a) presents the
temperature dependence of the longitudinal resistivity of sample
5B, demonstrates the clear zero field MIT, with a critical density
$p_c$$\approx$0.8 ($r_s$$\approx$36). Similar data and critical
density  were obtained on sample 5A. The critical point of MIT for
sample 11C and 11D is around $p_c$=0.95 ($r_s$$\approx$33),
reflecting the somewhat lower mobility of the wafer. All the $p_c$
values are lower than the critical density in
Ref.\onlinecite{HWJiangPRL1993,DCTsuiPhasediagram,HWJiangCvsBT},
attesting to stronger correlation effects in our samples.

The 2D hole density is determined by the dip positions of the
Shubnikov-de Haas (SdH) oscillations as $p$=$\nu$$B_\nu$e/h, where
$\nu$ is the Landau filling factor and $B_\nu$ is the
perpendicular magnetic field at the corresponding $\nu$. In
Fig.\ref{fig1}{b}, the longitudinal magneto-resistivity
$\rho_{xx}$($B$) for $p$=0.86 of sample QW 5B is shown at
temperatures from 50mK to 0.2K. At small magnetic fields, the
sample exhibits metallic behavior. However, it is striking to see
an unexpected insulating peak below 150mK between the $\nu$=1 and
the $\nu$=2 quantum Hall resistivity dips. This RIP is observed in
all of our four 10nm QWs. Another example of the RIP between
$\nu$=1 and $\nu$=2 QH states can be seen in Fig.1(c) of our
previous paper\cite{RichardandGao}. The resistance at the peak of
the RIP follows exponentially activated temperature dependence
(Fig.1C) and greatly exceeds the sample's zero field resistivity.
Apparently the temperature dependence of the insulating peak is
far stronger than the temperature dependence of the ordinary SdH
oscillation amplitude, which is determined by the thermal damping
factor, $X_T$/sinh($X_T$), where $X_T$=$2\pi^2k_BT$/$\hbar\omega$.
As another example, the peak resistivity for sample QW 5A exceeds
the value of 12K Ohms/square, six times larger than $\rho$ at
$B$=0T, at a temperature of 18mK and a density of $\emph{p}$=1.13
(Ref.\onlinecite{RichardandGao}). Fig.\ref{fig1}{c} illustrates
the temperature-dependent resistance of QW 11C at the peak of RIP
for densities from $\emph{p}$=1.21 to 0.9 in an Arrhenius plot.
Similar to the RIPs between fractional QH states in GaAs/AlGaAs 2D
electron\cite{VJGoldmann-GaAsreentrant} or
hole\cite{HWJiangp-GaAsreentrant,p-GaAsreentrant2005} systems, the
$R_{RIP}$ vs. $T$ of QW 5B and QW 11C is exponential,
$R_{RIP}$$\propto$exp($\Delta_{RIP}$/2T). The fittings according
to the thermal activation model give activation gaps
$\Delta_{RIP}$ up to 0.52K, shown in Fig.\ref{fig1}{d} for QW 5B
and QW 11C. For both QWs, the value of $\Delta_{RIP}$ appears to
increase exponentially as the density decreases. Besides, the
insulating phase emerges at higher density in the lower mobility
samples (QW 11C and 11D) than higher mobility samples (QW 5A and
5B), suggesting that disorder also contributes to the formation of
this RIP. The Hall resistance data\cite{RichardandGao,GaoHall}
indicate that the $\nu$=1 quantum Hall state is well formed but
Hall resistance is suppressed significantly ($\sim 20-30\%$) than
the classical value 1/$pe$ at the RIP and lower magnetic fields.

\begin{figure}[btph]
\centerline{\epsfig{file=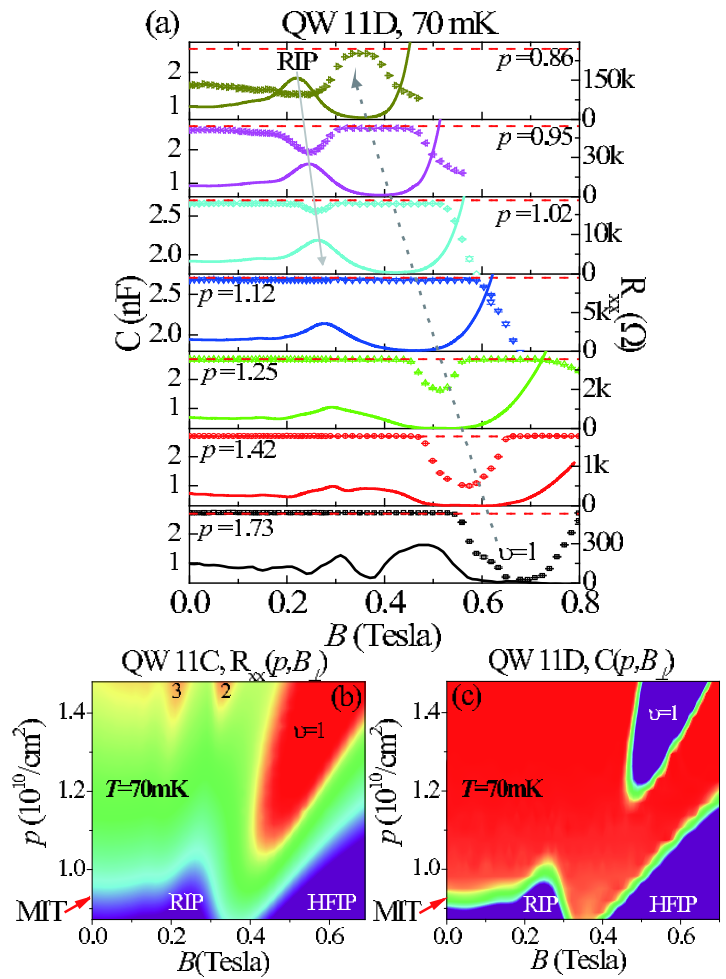,width=9.5cm}} \caption{(color
online) (a)Capacitance (symbol) and resistance (line) vs.
perpendicular magnetic field (0T-0.8T) for QW 11D at 70mK with
densities of $p$=0.86, 0.95, 1.02, 1.12, 1.25, 1.42 and 1.73. The
$\nu$=1 QH state gradually weakens as the density decreases, and
eventually disappears, followed by the emergence of the new RIP.
The dash line is the estimated value of geometric capacitance
$C_0$.(b)Longitudinal resistance map of QW 11C in $p$-$B_{\bot}$
plane at 70mK. As the color varies from red to violet, the
resistance increases from 300$\Omega$ up to 300$k\Omega$ in a log
scale. The area with resistance larger than 300$k\Omega$ is filled
with violet. (c)Capacitance of QW 11D in $p$-$B_{\bot}$ plane at
70mK. The capacitance drops from 2.7nF down to 2.2nF as the color
changes from red to violet. The area with capacitance less than
2.2nF is colored with violet.} \label{fig2}
\end{figure}

The similarity in the transport properties between the observed
insulating phase at $\nu>$1 and the previously studied RIPs
between fractional QH states suggests that the insulating phase
observed here is the RIP at $\nu$$\textgreater$1 in the phase
diagram proposed by Zhu and Louie (FIG.14 in
Ref.\onlinecite{XZhuMonteCarlo}(b)). Moreover, we performed
capacitance measurements to elucidate the thermodynamic
compressibility of the RIP observed here. Our experiment indicates
that this RIP is more $incompressible$ than both the $\nu$=1 QH
state and the $B$=0 state, offering further evidence supporting
the pinned Wigner crystal as the interpretation according to the
phase diagram in Ref.\onlinecite{XZhuMonteCarlo}(b).
Fig.\ref{fig2}{a} presents the magnetic-field-dependent top gate
capacitance (symbol) \emph{C(B)} for several densities ($p$=1.73,
1.42, 1.25, 1.12, 1.02, 0.95, 0.86) in QW 11D at 70mK, together
with the corresponding magneto-resistance curves (solid lines). In
the lower three panels where $p$$\textgreater$1.2, the capacitance
value drops strongly at the $\nu$=1 QH state, reflecting the 2D
system's small thermodynamic density of states (compressibility)
when the Fermi level resides in the Landau level energy gap at low
enough temperature. At low magnetic fields, the gate capacitance
remains constant around the geometric capacitance value (marked as
dashed horizontal lines in Fig.2a) due to the fact that quantum
capacitance $C_q$ of 2D systems is very large and does not
contribute much in the measured capacitance until strong Coulomb
interactions reduces the compressibility significantly in the low density regime. 
What is most interesting in Fig.2 is, however, the behavior of
capacitance as 2D hole density $p$ decreases towards the critical
density ($p_c\sim$0.95). As Fig.2a shows, the capacitance dip at
$\nu$=1 weakens as density decreases (dotted grey arrow). But a
new dip in the capacitance vs. $B$ curve emerges at the position
where transport data show the reentrant insulating behavior (light
grey arrow). It is remarkable to note that the compressibility of
the RIP is not only smaller than the zero field state, but also
the $\nu$=1 QH state. Moreover, since capacitance values smaller
than the geometric value was observed, the sign of the
compressibility for the RIP is $positive$, in contrast to the well
known negative compressibility of 2D electron liquids. In addition
to the capacitance dip at the RIP between $\nu$=1 and 2, the
sample's capacitance also drops at $\nu$$<$1, corresponding to the
HFIP.
\begin{figure*}[btph]
\centerline{\epsfig{figure=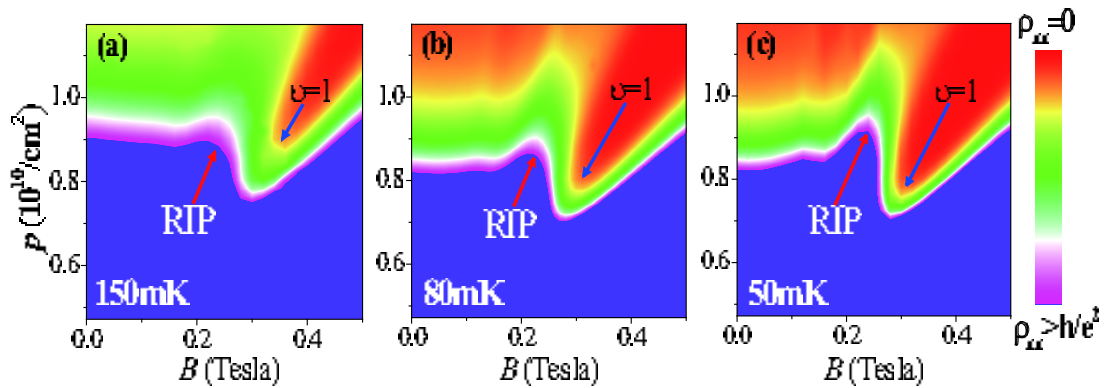,width=14.5cm}}
\caption{(color online) (a)(b)(c) Contour map of the longitudinal
resistivity of a 10nm wide p-type GaAs QW (sample QW 5B) in
$p$-$B_{\bot}$ plane at three temperatures: 150mK, 80mK, 50 mK.}
\label{fig3}
\end{figure*}

Previously, it is known that the 2D system's compressibility
approaches zero in the insulating phase of MIT\cite{HWJiangCvsBT}.
This behavior is also observed in our sample, as shown by the
reduced capacitance value of our sample at $B$=0 when the carrier
density is decreased to below $p_c\sim$0.95 (top two panels in
Fig.2a). But for hole densities around the critical density, the
application of perpendicular magnetic field induces RIP with
strongly insulating transport behavior as well as lower
capacitance (compressibility) than the zero field state. This
observation implies that the RIP is somehow related to the zero
field insulating phase of 2D MIT, yet perpendicular $B$ helps the
insulating phase forming more easily. To clarify the relation
between the new RIP and the zero MIT, we examine the 2D hole
system's phase diagram against $p$ and $B_{\bot}$ by both
transport and capacitance measurements. Figures \ref{fig2}{b} and
\ref{fig2}{c} plot the colored contour maps of longitudinal
resistance and capacitance for QW 11C and QW 11D from 0T up to
0.8T over the density range 0.8$\textless$$p$$\textless$1.5 at
70mK. In Fig.\ref{fig2}{b}, the three red finger-shaped areas
pointing towards origin are the $\nu$=1, 2, and 3 QH states.
Similarly, the violet finger in Fig.\ref{fig2}{c} represents
$\nu$=1 QH state. Due to the dominance of $C_0$ in measured $C$,
the $\nu$=2 and 3 QH states were not resolved in our capacitance
data. In Fig.\ref{fig2}{b} of transport map, the insulating phase
is shown in violet at the bottom, which also has low capacitance
(compressibility) as seen in Fig.\ref{fig2}{c}. Through comparing
the two maps, it is significant to see a unified phase boundary
linking zero field MIT with the new RIP and HFIP. First, the two
different methods give the same phase boundary of the MIT from the
$B$=0 limit up to $\nu$=1. Both maps indicate that the zero field
insulator is linked with the new RIP, suggesting a common origin.
It is necessary to point out that our phase diagram differs from
previous study\cite{DCTsuiPhasediagram} in that there is a RIP
protruding into the high density metallic regime while previous
finding in higher density samples showed a phase boundary line
moving down monotonically as $B_\perp$ increases towards the
$\nu$=1 QH \cite{DCTsuiPhasediagram}. Our phase boundary coincides
with the phase diagram in FIG.14 of
Ref.\onlinecite{XZhuMonteCarlo}(b), which predicts an insulating
solid phase between the $B$=0 liquid state and $\nu$=1 QH state.

We also studied the temperature evolution of the phase boundary of
the metal-insulator transition in the $p$-$B_{\bot}$ map via
transport measurement. Figure \ref{fig3} shows the longitudinal
resistance maps of QW 5B from $B$=0T to 0.5T and a density range
0.5$\textless$$p$$\textless$1.2 at 150mK, 80mK and 50mK. This
evolution depicts an interesting competition between the zero
field MIT and the formation of integer QH and the RIP. First, as
\emph{T} decreases, both the low field metallic state and $\nu$=1
QH become better formed as shown by the growing low resistivity
(red) areas in the phase diagram towards lower density. At the
same time, the RIP starts to form and protrude towards higher
density in the diagram. Eventually, at lowest $T$, there is a RIP
separating the low field metal and $\nu$=1 QH liquid for certain
density range above the critical density $p_c$=0.8. Then the phase
boundary becomes closer to the theoretical result predicted in
Ref.\cite{XZhuMonteCarlo}. Naturally, the implication is that the
new RIP and $\nu$=1 QH state are the ground states in the $T$=0K
limit. In addition, by comparing Fig.\ref{fig2}{b} and {c} with
Fig.\ref{fig3}, it is clear to see that the phase boundary shifts
to lower density as the sample mobility increases (less disorder).
Obviously, further study is necessary to understand better the
evolution of the liquid-WC in 2D as a function of $T$ and the role
of disorder.

The phase diagram of our \emph{narrow} QWs is distinct from those
of wide QWs or heterstructures
\cite{DCTsuiPhasediagram,HWJiangCvsBT,p-GaAsreentrant2005}. The
RIP between the $\nu$=1 and $\nu$=2 quantum Hall states in
strongly interacting 2D systems, even proposed decades
ago\cite{XZhuMonteCarlo}, has never been reported in literature.
We believe that the narrow QW width is a key in the formation of
this RIP at $\nu>$1. Empirically, it has already been noticed that
the RIP in GaAs 2D electron system shifts to higher filling factor
as QW width decreases \cite{shiftofRIP}. Another argument is in
GaAs hole system, Landau-level mixing is significant and the well
width will affect the hole effective mass, which is essential to
the Monte Carlo calculation \cite{XZhuMonteCarlo}. Furthermore,
spin-polarization was seen to induce more dramatic effect in
driving metallic state into insulating phase in narrow QW system
\cite{GaoparallelB}. Therefore, the strong spin-polarization close
to the $\nu$=1 QH state may have also helped the formation of the
RIP.


In summary, we have observed a new reentrant insulating phase
(RIP) between  the $\nu$=1 and $\nu$=2 QH states in a dilute
p-GaAs quantum well system with narrow width and found that this
RIP is connected with the zero field insulating phase of the 2D
metal-insulator transition. The RIP is similar to those observed
among the fractional QH states in other GaAs hetero-structures or
QWs with wider width. Based on transport and thermodynamic
compressibility experiments, we propose its origin to be related
to the formation of disorder pinned Wigner crystal of low density
2D holes. Both our transport and thermodynamic phase diagrams
match the liquid-Wigner crystal phase diagram proposed by Zhu and
Louie \cite{XZhuMonteCarlo}, leading to the suggestion that the
zero field MIT is a liquid-WC transition
\cite{microemulsion,microemulsion2,mitreview}.

The authors thank funding support from NSF (grant number
DMR-0906415). The work at Princeton was partially funded by the
Gordon and Betty Moore Foundation as well as the NSF MRSEC Program
through the Princeton Center for Complex Materials (DMR-0819860).

\end{document}